
\centerline{\bf Ashtekar variables, self-dual metrics, and $w_\infty$}
\bigskip
\centerline{Viqar Husain\footnote{$^*$}{email: viqar@itsictp.bitnet.
 Address after 30th September 1992: Dept. of Physics,
 University of Alberta, Edmonton, Canada T6G 2J1.}}
\centerline{International Center for Theoretical Physics}
\centerline{P. O. Box 586, 34100 Trieste, Italy }
\bigskip
\centerline{\bf Abstract}
\bigskip
\noindent
 The self-duality equations for the Riemann tensor are studied using the
 Ashtekar Hamiltonian formulation for general relativity. These equations may
be written as dynamical equations for three divergence free vector fields
 on a three dimensional surface in the spacetime.
\medskip\noindent
A simplified form of these equations, describing metrics with a one Killing
 field symmetry are written down, and it shown that a particular sector
of these equations has a Hamiltonian form where the
Hamiltonian is an arbitrary function on a two-surface. In particular,
   any element of the $w_\infty$ algebra  may be chosen as a the Hamiltonian.

 \medskip\noindent
For a special choice of this Hamiltonian, an infinite set of
solutions of the   self-duality  equations are given. These
 solutions are parametrized by elements of the $w_\infty$ algebra, which
in turn leads to an explicit form of four dimensional complex
 self-dual metrics
 that are in one to one correspondence with elements of this algebra.

 \vfill
\eject

\noindent {\bf Introduction}
\bigskip
\noindent Self-duality plays an important role in four dimensional field
theories that involve two-forms.
 In Yang-Mills theories the solutions to the Euclidean space field equations
 with self-dual curvature,
the `instantons', are solutions that give a finite action. Because
of this, these field
 configurations play an important role in the evaluation of
 the path integrals that involve Yang-Mills fields.   In the past few years
 these  solutions have played an essential role in mathematics as well,
 in the question of constructing invariants of four manifolds. It
 has been shown that certain invariants may  be calculated
from path integrals of `topological' field theories, theories
that are diffeomorphism invariant and have no local degrees of freedom [1].
\medskip
\noindent In general relativity, one role played by self duality is that
it provides a way of investigating a subset of the space of solutions
to  the Einstein equations. A substantial amount of work has been
done in this area and there are several interesting results.
It has been shown that for asymptotically flat spacetimes, the scattering
problem for self-dual initial data is trivial [2], that is, the evolved data at
future null infinity is the same as the ingoing selfdual data at past null
infinity.
Penrose has proposed that the `non-linear gravitons' [3],  metrics with
self-dual and anti-self-dual Weyl curvature, should play the essential role in
quantum gravity as the fundamental quanta,  as opposed to the linear spin two
particles.  A purely classical result of this work is that the general solution
of the self-dual equations may be written down. This solution is
 however in twistor space and the general solutions, as spacetime metrics, are
not available. Neither is there a Hamiltonian formulation for this sector
of the Einstein equations, and the integrability according to Liouville,
 which involves constructing an infinite set of commuting conserved
quantities, hasn't been demonstrated.
\medskip
\noindent In attempts at quantization of general relativity, it now appears
that self-duality  may again play an essential role [4]. The Hamiltonian
 formulation of the Einstein equations due to Ashtekar makes use of the
self-dual
 part of the spin-connection. Its projection onto spacelike surfaces
provides a
coordinate on the phase space of general relativity, and a spatial (densitized)
triad is the conjugate momentum.  These variables have proven to
be quite useful in pursuing the Dirac quantization procedure [5,6].
\medskip
\noindent Some connections between two dimensional conformal field theories and
 the self-dual Einstein equations have been discussed recently [7,8]. An
interesting
 result of this work is that the self-duality equations for a class
of metrics are the same as the field equations for a particular two dimensional
conformal field theory, the continuum limit of the Toda theory. Using this
 connection it was demonstrated in [7] that the self-dual Einstein equations
 for this class of metrics have an associated infinite dimensional symmetry
algebra,
 the algebra of area preserving diffeomorphisms of a two dimensional
surface [9], a subset of which is the $w_\infty$ algebra.  This algebra
 is the generalization of the Virasoro algebra that includes all higher
spins 2,3,...$\infty$. Below, this result
 appears in a direct way where it is shown
that a sector of the self-duality equations has a Hamiltonian form
 where the Hamiltonian is an arbitrary function on a two dimensional
surface, and furthermore, that metrics parametrized by elements of the
 $w_\infty$ algebra may be explicitly written down.

 \medskip
\noindent For the purposes of this paper, the Ashtekar hamiltonian variables,
being   connected  with self-duality,
  provide a  simple way to write down a first order form of the (anti)self-dual
Einstein equations [10].
 It has been shown that this form of the self-dual Einstein equations may
 be derived from the self-dual Yang-Mills equations [11], and they also
appear as half of the Hamilton equations in the weak coupling
 ($G\rightarrow 0$) limit  of the full Einstein equations (written in the
Ashtekar formulation) [12].
  \medskip
\noindent The contents of this paper are as follows: in the next section, the
derivation of the self-dual Einstein equations from  the Ashtekar variables
 is reviewed. In section 3 these equations
are simplified via a one Killing field reduction, and it is shown that
 a sector of the resulting equations has a natural Hamiltonian formulation
 where the Hamiltonian is an {\it arbitrary function} on a
two-surface. The Hamiltonians may be chosen as elements of the $w_\infty$
 algebra, and for a particular such choice, the Hamilton equations are
integrated explicitly. The initial data used for this integration has a
 $w_\infty$ algebra associated with it, and the resulting  metrics are
 therefore in one to one correspondence with elements of this algebra. The
 metrics are given explicitly.
 The last section
is a discussion of the results regarding connections with two-dimesional
physics and possibilities for constructing more general solutions, as well as
 quantization.
\bigskip
\noindent {\bf The self-dual equations}
\bigskip
\noindent This section is a review of the derivation of the self-dual Einstein
equations
from the Ashtekar hamiltonian variables [10].
\medskip
\noindent The Einstein equations for real
metrics on a real Euclidean manifold may be derived from a Palatini type action
 where the field variables are  vierbeins $e_\mu^i $, and the
 spin-connection $\omega_\mu^{ij}$. The $\mu,\nu...=0,...,3$ are spacetime
indices and
$i,j...=0,...,3$ are internal so(4) indices. The lagrangian density is the
 4-form  $\epsilon_{ijkl}(e^i\wedge e^j\wedge R^{kl}[\omega])$, where $R^{kl}$
is the curvature 2-form of $\omega^{ij}$. The Hamiltonian 3+1 decomposition of
this action leads essentially to the usual form of the phase space and
constraints [13].
 \medskip
\noindent An alternative lagrangian density [14] that
 also leads to the Einstein equations is formed by replacing the curvature of
 $\omega^{ij}$ by the curvature of its (anti)self-dual part with respect to
 the internal so(4) indices:$^{\pm}A^{ij}\equiv {1\over 2}(\omega^{ij}
\pm{1\over 2}\epsilon^{ij}_{\ \ kl}\omega^{kl})$.
  The new action, if we use the anti-self-dual part $\ ^-A^{ij}$ is
$$S=\int_M \epsilon_{ijkl}\ e^i\wedge e^j\wedge\ ^{-}F^{kl}[^{-}A]\eqno{(0)}$$
where $^-F^{ij}$ is the curvature of $^-A^{ij}$.
  This replacement does not affect the Einstein equations
 because $^{-}F ^{ij}= {1\over 2}(R^{ij} - {1\over 2}
\epsilon^{ij}_{\ \ kl}R^{kl})$, and the variation of the second term
as a function of the vierbein is zero.
(Recall that
 $R^{ij}[\omega] = R^{ij}[^-A \ +\ ^+A] \equiv \ ^-F^{ij}[^-A]\ +
\ ^+F^{ij}[^+A]$). It is the Hamiltonian decomposition of the action (0)
that leads to the Ashtekar variables.
(Note that
 one can use either the self-dual or the anti-self-dual part of the
curvature,
 (or equivalently, the connection), to construct the Lagrangian.
We are using the anti-self-dual part in order to get equations for self-dual
metrics, as will become clear below).
\medskip
\noindent To find the Hamiltonian versions of such actions on (Euclidean)
 4-manifolds $M$, one assumes the spacetime is
 of topology $\Sigma \times R$. $R$ is the `time' direction, which is
fixed by introducing a vector field $t^\mu$ on $M$ such
that $t^\mu=Nn^\mu$, where $n^\mu$ are unit normals to the `spacelike'
surfaces $\Sigma$ and $N$ is the lapse {\it function}.
Each surface carries a `time' label $t$. The function $t$
 is determined by ${\cal L}_t  t = 1$, where ${\cal L}_t$ denotes the Lie
derivative with respect to $t^\mu$. The $n^\mu$ may be used determine the
metric on $\Sigma$, $q$, from the spacetime metric $g$ :
$q_{\mu\nu }= g_{\mu\nu} -n_\mu n_\nu$ (with $g_{\mu\nu}$ determined by
the vierbeins). This `spatial' metric may be
used to project tensors on $M$ to tensors on $\Sigma$.
\medskip
\noindent   For the action (0), the phase space
 coordinate is the spatial projection of $^{-}A_\mu^{ij}$, $A_a^i$,
which is now valued in so(3) ($i=1,..,3$),
 and the conjugate momentum is the densitized dreibein $E^a_i$.
($a,b...=1,..,3$ denote spatial indices). The fundamental Poisson bracket
is
 $$ \{A_a^i(x),E^{bj}(y)\}=\delta^{ij}\delta^b_a\delta^3(x-y) $$
 (For details of the Hamiltonian decomposition see [5,6,13,14]).
The first class constraints on the phase space
 corresponding to the invariances of the action are
$$ \partial_a E^{ai} + \epsilon^{ijk} A_a^jE^{ak} = 0  \eqno{(1)}$$
$$ F_{ab}^iE^{bi} = 0 \eqno{(2)} $$
$$ H\equiv \epsilon^{ijk}E^{ai}E^{bj}F_{ab}^k = 0 \eqno{(3)}$$
\noindent where $F_{ab}^i = \partial_{[a}A_{b]}^i +
\epsilon^{ijk}A_a^jA_b^k$.
 (1) is the Gauss law constraint generating triad rotations
and gauge tranformations, (2) generates spatial diffeomorphisms, and
 (3) generates time reparametrizations or `evolution' of the initial data on
the spacelike surfaces. The evolution equations are the Hamilton equations
 for the phase space variables with respect to an arbitrary
 `lapse' {\it density}  ${\cal N}$, which is a density of weight $-1$:
$H({\cal N})=\int_\Sigma {\cal N}H$ and

 $$  \dot{E}^{ai} = \{ E^{ai}, H({\cal N}) \}_{PB} =
 \epsilon^{ijk}D_b({\cal N}E^{aj}E^{bk})\eqno{(4)}$$
\noindent and
 $$ \dot{A}_a^i = \{ A_a^i , H({\cal N}) \}_{PB}=
{\cal N}\epsilon^{ijk}E^{bj}F_{ab}^k \eqno{(5)}$$
\noindent where $D_a$ is the covariant derivative for the connection $A_a^i$.
Note that ${\cal N}$ is related to the lapse function $N$ by
 $N=(det q)^{1/ 2} {\cal N}$.
 The equations (4-5), together with the constraints (1-3), are
equivalent to the full empty space Einstein equations without a cosmological
 constant.
\medskip
\noindent We now consider the restrictions of these that will yield
(Euclidean)   spacetime metrics $g_{\mu\nu} $ with self-dual Riemann tensor:
$R_{\mu\nu\alpha\beta}
 ={1\over 2}\epsilon_{\mu\nu}^{\ \ \gamma\delta}R_{\gamma\delta\alpha\beta}$.
 That this  is equivalent to the condition $R_{\mu\nu}^{ij} = {1\over 2}
 \epsilon^{ij}_{\ \ kl} R_{\mu\nu}^{kl} $ is easily verified (using
 $R_{\mu\nu\alpha\beta} = R_{\mu\nu}^{ij}e_{\alpha i}e_{\beta j}$).
The  self duality condition is (by definition)
 equivalent    to the condition $^-F_{\mu\nu}[^-A]=0$. Now, since the spatial
 projection of this $^-F$ is determined  by the spatial metrics $q$ via
 $ F_{\mu\nu} = q_\mu^\alpha q_\nu^\beta (\  ^-F_{\alpha\beta})$,
this implies that the curvature of the phase space coordinate $A_a^i$,
$F_{ab}^i[A]$, is zero if $^-F_{\alpha\beta}=0$.
 Thus, $F_{ab}^i=0$ is the phase space condition corresponding to the
self-duality of the spacetime Riemann curvature.  One can now work in a gauge
where $A_a^i$ itself is zero. Either
 of these conditions  immediately solve the constraints (2) and (3).
 Also, if $A_a^i$ is chosen to be zero on the initial data surface,
it remains
zero under the evolution equations (5). The remaining equations,
with this condition imposed are
$$ \partial_a E^{ai}=0 \eqno{(6)}$$
$$  \dot{E}^{ai} =  \epsilon^{ijk}\partial_b({\cal N}E^{aj}E^{bk})
\eqno{(7)}$$

\noindent These equations may be further  simplified my making a specific
 choice for the arbitrary lapse density ${\cal N}$. This corresponds to fixing
`time' gauge. Choosing ${\cal N}$ to be a constant
 $(\dot{\cal N}\equiv{\cal L}_t {\cal N}=0;\ \partial_a{\cal N}=0)$ and
defining the vector fields
 $V^a_i= {\cal N}E^a_i$ (recall that ${\cal N}$ and $E^a_i$ are densities of
weight $-1$ and $+1$ respectively),  we can rewrite these remaining equations
 as equations for the three vector fields $V^a_i$
 $$   \partial_a V^a_i=0    \eqno{(8)}$$
$$     \dot{V}^a_i=\epsilon_{ijk}[V_j,V_k]^a  \eqno{(9)}$$
 where [\ ,\ ] denotes the Lie bracket of the vector fields. Equations (8-9)
are the new form of the self-dual Einstein equations. The spacetime metric
is constructed
 from  a solution of these equations via
$$ g^{ab} = D^{-1}{\cal N}(V^a_iV^b_i +t^at^b) \eqno {(10)} $$
where $D=$ det$V = \epsilon^{ijk}\epsilon_{abc}V^a_iV^b_jV^c_k$.
\medskip
\noindent
Although real so(4) connections and vierbeins were used in the steps
leading to equations (8-9), which give real Euclidean metrics, the complex
metric case may be discussed similarly. The same final equations result.
Therefore, complex   metrics with self-dual Riemann curvature may be
constructed
 by starting with  a triad of complex vector fields $V_i^a$ satisfying
(8-9).
 \bigskip
\noindent {\bf The self-dual metrics}
\bigskip
\noindent In this section, a one Killing field reduction of the self-duality
 equations (8) and (9) is described. It is shown that a sector of these
equations has a  Hamiltonian form where the Hamiltonians
 are arbitrary functions on a two-surface. For a specific Hamiltonian, an
infinite class of solutions to the Hamilton equations are given. These
solutions may be associated with elements of the $w_\infty$ algebra and the
infinte set of metrics determined by them are presented.

 \medskip
\noindent  The spatial surface $\Sigma$ has a fixed volume element
$\tilde{\Omega}_{abc} $. This is because the divergence free condition
 (8) is with respect to a fixed non dynamical
  volume form. We fix a flat coordinate system $(x,y,\theta)$
 on some neighbourhood of $\Sigma$.
  We would like to construct  a self-dual metric that  has a one Killing
 vector field symmetry with respect to the vector field $u^a\equiv
(\partial/\partial\theta)^a$. $\Sigma$ will then have the topology
$R\times\Sigma^2$ for
arbitrary two surfaces $\Sigma^2$, on which the coordinates are $(x,y)$.
(It may
be possible to compactify the orbits of the Killing field to $S^1$ after
specific metrics are determined).
 A two dimensional  volume form $\Omega$ on $\Sigma^2$ can be fixed by
$$ \Omega_{ab}=\tilde{\Omega}_{abc} ({\partial\over \partial\theta})^c
 \eqno{(11)} $$
Locally, the coordinates $(x,y)$ may be chosen such that
$\Omega_{xy}=-\Omega_{yx}=1$. This volume form can be used to define
 Poisson brackets for functions on (a local neighbourhood) of $\Sigma^2$ via
$\{f,g\}=\Omega^{ab}\partial_af\partial_bg$.  These brackets will be  used
below.
\medskip
\noindent In order to satisfy the Killing symmetry condition ${\cal L}_u
g_{ab}=0$, we
must have by (10) that $[u,V_i]^a=0=[u,t]^a$ for all $i$. An ansatz for
$V^a_i$
that satisfies this condition and solves the divergence free conditions
(8) is
 $$ V_i^a = \Omega^{ab}\partial_b\Lambda_i + \Pi_i
({\partial\over \partial\theta})^a  \eqno{(12)} $$
 where $\Lambda_i$ and $\Pi_i$ are at this stage six arbitrary real functions
of  $(x,y,t)$. Substituting these into the dynamical equations (9)
 gives in a straightforward way the equations for these
functions.
 $$  \dot{\Lambda}_i=
 \epsilon_{ijk}\{\Lambda_k,\Lambda_j\} \eqno{(13)} $$
and
$$\dot{\Pi}_i = 2
\epsilon_{ijk}\{\Pi_k,\Lambda_j\}  \eqno{(14)}$$

\medskip
\noindent
We now consider a particular set of solutions to these equations.  This may be
viewed as an ansatz to solve (13-14). Set  $\Lambda_1=\Lambda (x,y,t)$,
 $\Lambda_2=-i\Lambda (x,y,t)$, and $\alpha =\Pi_1+i\Pi_2$. The first two
 of these
imply that $\Lambda_3$ is arbitrary and time independent, and equations (13-14)
become
  $$ \dot{\Lambda} = \{ \Lambda, H\} \eqno{(15a)}  $$
  $$ \dot{\alpha} = \{\alpha,H\} \eqno{(15b)} $$
 $$ \dot{\Pi}_3 = 2i \{\alpha^*,\Lambda\}   \eqno{(15c)}  $$
  $$ \{\Pi_3,\Lambda\}=0    \eqno{(15d)} $$
 where * denotes complex conjugation. (Note that the choice of $\Lambda_2$
makes the vector fields (12) complex and so the metrics (10) will also be
 complex). (15a) and (15b) are Hamiltonian evolution
 equations for $\Lambda$ and $\alpha$ with arbitrary  Hamiltonian
 $H\equiv 2i\Lambda_3$, and (15c)-(15d) determine $\Pi_3$ via an evolution and
 a `constraint' equation from solutions of the first two. The constraint
equation results from the consistency condition that the evolution equations
 for $\alpha$  and $\alpha^*$  be complex conjugates of each other.

 \medskip
\noindent
There is a natural way to write down a series of Hamiltonians  $H$ (or choices
of the  functions $\Lambda_3$), that
 form a $w_\infty$ algebra. The functions $w_n^s(x,y) = x^{n+s-1}y^{s-1}$
 form this algebra via Poisson brackets with respect to (11):
 $$ \{w^s_m,w^{t}_n\}=((t-1)m - (s-1)n)w_{m+n}^{s+t-2} \eqno{(16)}$$
where $m,n\in Z$ and $s,t$ are integers $\ge 2$. This algebra
contains the Virasoro algebra which arises for $s=t=2$
 $$ \{w^2_m,w^2_n\}=(m - n)w_{m+n}^2. \eqno{(17)}$$
The $w_n^s$ also satisfy the relation
$$ \{w_m^2,w_n^s\}=((s-1)m-n)w_{m+n}^s \eqno{(18)} $$
\medskip
\noindent
In order to find some explicit solutions to (15), we consider the specific
Hamiltonian
$ H = i w^2_0 = ixy$ ($\Lambda_3 =  w^2_0/2)$ and the initial condition
$\Lambda (x,y,t=0)=w^s_m = x^{m+s-1}y^{s-1}$. Then it is easy to verify using
(18) that some solutions to  (15a) are
$$ \Lambda_m^s (x,y,t) = \Lambda_0e^{imt}w^s_m (x,y) \eqno{(19)} $$
where $\Lambda_0$ is an arbitrary constant.
These solutions have a natural associated $w_\infty$ symmetry namely
$$ \{\Lambda_m^s,\Lambda_n^t\}= ((t-1)m-(s-1)n)\Lambda_{m+n}^{s+t-2} $$
 which is just the (built in) symmetry associated with the initial data
$\Lambda(x,y,t=0)$.
Note also that the Hamiltonian used to obtain these solutions is in fact
just $L_0 (\equiv w^2_0)$ of the Virasoro algebra (17)
 (where $L_n\equiv w_n^2$).

\medskip
\noindent
Given these solutions to ($15a$), one still needs to solve (15b)-(15d) to
 determine
 metrics explicitly. One easy solution is to set the $\Pi_i$ equal to three
constants $C_i$. With this the vector fields (12) become
 $$ V^a_1=\Lambda_0e^{imt}\Omega^{ab}\partial_b w^s_m +
C_1({\partial\over \partial\theta})^a $$
$$ V^a_2=-i\Lambda_0e^{imt}\Omega^{ab}\partial_b w^s_m +
C_2({\partial\over \partial\theta})^a $$
$$ V^a_3={1\over 2}\Omega^{ab}\partial_b w^2_0+
C_3({\partial\over \partial\theta})^a \eqno{(20)} $$
The metric corresponding to this solution obtained from
 (10) is
$$ ds^2 = -2i(Ax+By)^{-1}[-{1\over 4}(Ax+By)^2dt^2 + ({Cy^2\over
4}-A^2)dx^2 +  ({Cx^2\over 4}-B^2)dy^2$$
$$ -{1\over 2}(Axy+By^2)dxd\theta - {1\over 2}(Ax^2+Bxy)dyd\theta]
\eqno{(21)} $$
where
$$ A= -\Lambda_0e^{imt}(C_1-iC_2)\partial_x w^s_m - {C_3y\over 2} \ \ \ \
\ \  B=\Lambda_0e^{imt}(C_1-iC_2)\partial_y w^s_m + {C_3x\over 2} \ \ \ \ \ \
C = C_1^2+C_2^2+C_3^2
$$
We note that by letting $s=2$ in (21) we get a smaller set of metrics
 parametrized by elements of the Virasoro sub-algebra (17) of $w_\infty$.

\medskip
\noindent
  To find a more general solution to (15b)-(15d), we can write a
  a solution similar to (19) for $\alpha$:
$\alpha^{s^\prime}_n=\alpha_0e^{int}w_n^{s^\prime}$,
  where at this stage there is no relation
between $n,s^\prime$ here and the $m,s$ of (19). With this $\alpha$ we have
 from (15c) that $\dot{\Pi}_3$ $=2i\{w^{s^\prime}_n, w^s_m\}e^{i(m-n)t}$,
 and  from
 (15d) that $\{\Pi_3,\Lambda\} =
\Lambda_0 e^{imt}\{\Pi_3,w_m^s\}=0 $.
  The first of these gives the $(x,y)$ dependence of $\Pi_3$ to be
 proportional to $w^{s+s^\prime-2}_{m+n}$, and the second
  gives a relation between $s,s^\prime$ and $m,n$ given by
$$ \{w^{s+s^\prime-2}_{m+n},w_m^s\} =
[(s-1)(m+n)-(s^\prime + s-3)m  ]w^{2s + s^\prime -4}_{2m+n}=0$$
This is satisfied by $s^\prime = s+1$ and $m=n$. Thus, the solutions to
(15b)-(15d) consistent with (19) are
$$ \alpha=\alpha_0 e^{imt}w^{s+1}_m
\ \ \ \ \ \  \Pi_3 = -2im\alpha_0\Lambda_0 w^{2s-1}_{2m} t + k. \eqno{(22)} $$
where $k$ is an arbitrary constant.
The more general metrics obtained from this solution are obtained from (21)
by replacing the constants $C_i$ by the functions
$C_1 = \alpha_0{\rm cos}mt w_m^{s+1}$,
 $C_2=\alpha_0{\rm sin}mt w_m^{s+1}$, and $C_3=\Pi_3$.
\medskip\noindent
To summarize, the main result of this section are the metrics (21) labelled
by elements $w_m^s$ of the algebra (16), which are obtained by
using $L_0\equiv w_0^2$ as the Hamiltonian in equations (15). This result
displays rather explicitly the previously discussed [7] $w_\infty$ infinite
spin symmetries associated with the  self-duality condition of the
Riemann curvature.

\bigskip
\noindent{\bf Discussion}
\bigskip
\noindent
Using the Ashtekar variables, we have given an infinite set of metrics
 with self-dual Riemann curvature, where the metrics are parametrized by
elements of the $w_\infty$ algebra. This is however not an exhaustive
set of solutions for the one-Killing field reduction that is used. This
is because we have considered only one class of initial data for the
evolution equations and only one specific `Hamiltonian', $w_0^2$, in equations
 (15). One may choose other Hamiltonians and study
these equations further and it would be of interest to see what other
choices give a rich class of solutions.

 \medskip
\noindent
There are a number of known metrics with self-dual Riemann curvature [15]. If
the metrics are  real and geodesically complete without singularities,
 they are called gravitational instantons.
 Examples are  the Eguchi-Hanson and Taub-NUT metrics [15]. Such metrics are
determined
 by starting with an appropriate ansatz in some local coordinates,
 (such as spherical coordinates on $R^4$  for the Eguchi-Hanson case),
 and then determining the manifold globally by suitably eliminating any
 curvature
 singularities, or by determining maximal extensions in appropriate
coordinates, or both. In this regard it  would be interesting to
study the global structure of the metrics given here to see if they have
real sections that are free of singularities.
\medskip
\noindent
 There is an ansatz more general than (12) which involves adding
terms corresponding to elements of the cohomology  group, $H^1(\Sigma^2)$,
 if
for example,  one starts with $\Sigma^2=R^2$ with some punctures. The terms to
 be added to (12)  would be of the
form $g_r(t)\Omega^{ab}\omega_b^{(r)}$ for elements $\omega^{(r)}$
 $(r=1,..,n)$ of  $H^1(\Sigma^2)$ and arbitrary functions $g_r(t)$.
 More generally, one may try an ansatz using triads with specific symmetries
 and try to fix the spatial topologies from the start. In such cases also
 it would be relevant to consider the contributions discussed in this
 paragraph. Fixing the spatial topology by a suitable choice of triads may,
 however, lead to  a specific metric rather than a
large class  of metrics (as happens for example, with the Eguchi-Hanson
 ansatz).  It may still be useful to pursue this case since the $w_\infty$
 generators may be expanded is a basis geared to the appropriate
topology of the 2-surfaces, and one may get a similar
structure associated with solutions of (15) for more general topologies.
\medskip\noindent
 The reduction of the Einstein equations discussed here, involving
self-duality, a Killing symmetry and the ansatz that gives the
Hamiltonian form (15)  may be viewed as a particular
  midi-superspace model. Since there is an infinite
 dimensional symmetry algebra on the solution space, it may be possible to
describe a quantization in terms of representations of $w_\infty$.

\medskip
\noindent
This work was motivated by the desire to find a general solution to
the unreduced self-dual equations (8)-(9), and to
 construct explicitly the infinite number of associated conserved quantities.
This would, if it can be done, perhaps provide a connection in terms of
 metrics, with the
general twistor space solution to the self-dual equations [3].
\medskip
\noindent
I would like to thank P. Hajicek, D. Mazzitelli, K. S. Narain,
 S. Randjbar-Daemi and C. Rovelli
for helpful conversations, and ICTP for hospitality and financial support.

\bigskip
\noindent{\bf References}
 \bigskip

\noindent[1] E. Witten, Commun. Math. Phys. 17, 353 (1988).
 \smallskip

\noindent[2] M. Ludvigsen, E.T. Newman, K.P. Tod, Phys. Rept. 71, 51 (1981).
\smallskip

\noindent[3] R. Penrose, Gen. Rel. Grav. 7, 31 (1976).
\smallskip

\noindent[4] A. Ashtekar, Phys. Rev. Lett. 57, 2244 (1986); Phys. Rev. D36,
 1587 (1987).
\smallskip

\noindent[5] C. Rovelli, Class. Quant. Grav.  8, 1613 (1991).
\smallskip

\noindent[6] A. Ashtekar,  {\it Non-perturbative canonical gravity},
 Lecture
notes in collaboration with R. S. Tate (World Scientific, Singapore, 1991).
\smallskip

\noindent[7] Q. H. Park,  Phys. Lett. B236, 429, (1990); Phys. Lett. B238,
 287 (1990).
\smallskip

\noindent[8] I. Bakas and E. B. Kiritsis, in {\it Topological methods in
 quantum field theory},
ed. W. Nahm, S. Randjbar-Daemi, E. Sezgin and E. Witten,
 (World Scientific, Singapore, 1991).
\smallskip

\noindent[9] For a recent review see E. Sezgin, Preprint Texas A\&M, (1992).
\smallskip

\noindent[10] A. Ashtekar, T. Jacobson, L. Smolin, Comm. Math. Phys. 115, 631
 (1988).
\smallskip

\noindent[11] L. Mason, E. T. Newman, Comm. Math. Phys. 121, 659 (1989).
\smallskip

\noindent[12] L. Smolin,  Class and Quant. Gravity 9, 883 (1992).
\smallskip

\noindent[13] See the appendix in A. Ashtekar, A. P. Balachandran, and S. Jo,
Inter. Jour. Mod. Phys.  A4, 1493 (1989).
\smallskip

\noindent[14] J. Samuel, Pramana J. Phys. 28 L429 (1987); T. Jacobson and
 L. Smolin, Phys. Lett. B196, 39 (1987); Class. Quant. Grav. 5 (1988) 583.
\smallskip

\noindent[15] See for example the list in T. Eguchi, P.B. Gilkey, and
 A. J. Hansen, Physics Repts. 66, 213 (1980).
\smallskip

\end